\documentclass[rmp,aps,lengthcheck,nofootinbib]{revtex4}
\usepackage{longtable}
\usepackage{dcolumn}
\usepackage{epsf}
\usepackage{bm}
\usepackage{amsmath}
\usepackage{graphicx}
\usepackage{xr}
\usepackage{afterpage}
\begin{document}
%
\def\elabel#1{\label{#1}}
\def\plabel#1{\label{#1}}
\def\slabel#1{\label{#1}}
\def\tlabel#1{\label{#1}}
\def\flabel#1{\label{#1}}
\def\numcite#1{\cite{#1}}
\def\numtextcite#1{\textcite{#1}}
\def\lref#1{\ref{#1}}
\def\comment#1{}
%
\newcommand{\greeksym}[1]{{\usefont{U}{psy}{m}{n}#1}}
\newcommand{\rmssmu}{\mbox{\scriptsize{\greeksym{m}}}}
\newcommand{\rmsstau}{\mbox{\scriptsize{\greeksym{t}}}}
\newcommand{\rmssgamma}{\mbox{\scriptsize{\greeksym{g}}}}
\newcommand{\rmmu}{\mbox{\greeksym{m}}}
\newcommand{\rmtau}{\mbox{\greeksym{t}}}
\newcommand{\rmalpha}{\mbox{\greeksym{a}}}
\newcommand{\rmssalpha}{\mbox{\scriptsize{\greeksym{a}}}}
\newcommand{\rmpi}{\mbox{\greeksym{p}}}
\newcommand{\rmsspi}{\mbox{\scriptsize{\greeksym{p}}}}
\newcommand{\rmphi}{\mbox{\greeksym{f}}}
\newcommand\T{\rule{0pt}{2.4ex}}
\newcommand\B{\rule[-0.8ex]{0pt}{0pt}}
\def\lbar{\lambda\hskip-4.5pt\vrule height4.6pt depth-4.3pt width4pt}
\def\moles{$^{\rm o}\hskip-4.85pt\vrule height4.98pt depth-4.89pt width4.4pt\hskip4.85pt$}
\def\fr#1#2{{\textstyle{#1\over#2}}}
\def\nothing#1{\phantom{#1}}
\def\iG{\it \Gamma}
\title{
       CODATA Recommended Values of the Fundamental Physical
       Constants: 2014\footnote{
  This report was prepared by the authors under the auspices of 
  the CODATA Task Group on Fundamental Constants.
  The members of the task group are: \\
  F. Cabiati, Istituto Nazionale di Ricerca Metrologica, Italy \\
  J. Fischer, Physikalisch-Technische Bundesanstalt, Germany \\
  J. Flowers (deceased), National Physical Laboratory, United Kingdom \\
  K. Fujii, National Metrology Institute of Japan, Japan \\
  S. G. Karshenboim, Pulkovo Observatory, Russian Federation \\
  E. de Mirand\'es, Bureau international des poids et mesures \\
  P. J. Mohr, National Institute of Standards and Technology, United States of America \\
  D. B. Newell, National Institute of Standards and Technology, United States of America \\
  F. Nez, Laboratoire Kastler-Brossel, France \\
  K. Pachucki, University of Warsaw, Poland \\
  T. J. Quinn, Bureau international des poids et mesures \\
  C. Thomas, Bureau international des poids et mesures \\
  B. N. Taylor, National Institute of Standards and Technology, United States of America \\
  B. M. Wood, National Research Council, Canada \\
  Z. Zhang, National Institute of Metrology, China (People's Republic of) \\
  }
  }
\author{
Peter J. Mohr\footnote{Electronic address: mohr@nist.gov},
David B. Newell\footnote{Electronic address: dnewell@nist.gov},
Barry N. Taylor\footnote{Electronic address: barry.taylor@nist.gov}
}
\affiliation{
    National Institute of Standards and Technology,
    Gaithersburg, Maryland 20899-8420, USA
   }

\date{\today}

\begin{abstract}

This document gives the 2014 self-consistent set of values of the constants
and conversion factors of physics and chemistry recommended by the
Committee on Data for Science and Technology (CODATA).  These values are
based on a least-squares adjustment that takes into account all data
available up to 31 December 2014.  The recommended values may
also be found at physics.nist.gov/constants.  \\

\end{abstract}
\maketitle

\newpage

\def\s#1{\hbox to #1pt{}}
\def\hsp{\hbox to 15 pt{}}
\def\b{\hbox to 10 pt{}}
\begin{table*}[h]
\caption{An abbreviated list of the CODATA recommended values of the
fundamental constants of physics and chemistry based on the 
2014 adjustment.}
\tlabel{tab:abbr}
\begin{tabular}{l@{\hsp}l@{\hsp}l@{\hsp}l@{\hsp}l}
\toprule
& & & & Relative std. \\
\s{35}Quantity & \s{-10}Symbol & \s{15}Numerical value & \s{2}Unit 
& uncert. $u_{\rm r}$ \\
\colrule
speed of light in vacuum & $ c,c_0 $ & 299\,792\,458 & m~s$^{-1}$ & exact
\vbox to 12 pt {} \\
magnetic constant & $\mu_0$ & $ 4\rmpi\times10^{-7}$ & N~A$^{-2}$ & \\
& & $=12.566\,370\,614...\times10^{-7}$ & N~A$^{-2}$ & exact \\
electric constant 1/$\mu_0c^{2}$ & $\epsilon_0$ &
$8.854\,187\,817...\times 10^{-12}$ & F~m$^{-1}$ & exact \\
Newtonian constant
of gravitation~~ & $ G $ & $ 6.674\,08(31)\times 10^{-11}$ & m$^{3}$~kg$^{-1}$~s$^{-2}$ & $ 4.7\times 10^{-5}$ \\
Planck constant & $ h $ & $ 6.626\,070\,040(81)\times 10^{-34}$ & J~s & $ 1.2\times 10^{-8}$ \\
\b $h/2\rmpi$ & $\hbar$ & $ 1.054\,571\,800(13)\times 10^{-34}$ & J~s & $ 1.2\times 10^{-8}$ \\
elementary charge & $ e $ & $ 1.602\,176\,6208(98)\times 10^{-19}$ & C & $ 6.1\times 10^{-9}$ \\
magnetic flux quantum $h$/2$e$ & ${\it \Phi}_0$ & $ 2.067\,833\,831(13)\times 10^{-15}$ & Wb & $ 6.1\times 10^{-9}$ \\
conductance quantum $2e^2\!/h$ & $G_0$ & $ 7.748\,091\,7310(18)\times 10^{-5}$ & S & $ 2.3\times 10^{-10}$ \\
electron mass & $ m_{\rm e}$ & $ 9.109\,383\,56(11)\times 10^{-31}$ & kg & $ 1.2\times 10^{-8}$ \\
proton mass & $ m_{\rm p}$ & $ 1.672\,621\,898(21)\times 10^{-27}$ & kg & $ 1.2\times 10^{-8}$ \\
proton-electron mass ratio & $m_{\rm p}$/$m_{\rm e}$ & $ 1836.152\,673\,89(17)$ & & $ 9.5\times 10^{-11}$ \\
fine-structure constant $e^2\!/4\rmpi\epsilon_0 \hbar c$ & $\alpha$ & $ 7.297\,352\,5664(17)\times 10^{-3}$ & & $ 2.3\times 10^{-10}$ \\
\b inverse fine-structure constant & $\alpha^{-1}$ & $ 137.035\,999\,139(31)$ & & $ 2.3\times 10^{-10}$ \\
Rydberg constant $\alpha^2m_{\rm e}c/2h$ & $ R_\infty$ & $ 10\,973\,731.568\,508(65)$ & m$^{-1}$ & $ 5.9\times 10^{-12}$ \\
Avogadro constant & $N_{\rm A},L$ & $ 6.022\,140\,857(74)\times 10^{23}$ & mol$^{-1}$ & $ 1.2\times 10^{-8}$ \\
Faraday constant $N_{\rm A}e$ & $ F $ & $ 96\,485.332\,89(59)$ & C~mol$^{-1}$ & $ 6.2\times 10^{-9}$ \\
molar gas constant & $ R $ & $ 8.314\,4598(48)$ & J~mol$^{-1}$~K$^{-1}$ & $ 5.7\times 10^{-7}$ \\
Boltzmann constant $R$/$N_{\rm A}$ & $k$ & $ 1.380\,648\,52(79)\times 10^{-23}$ & J~K$^{-1}$ & $ 5.7\times 10^{-7}$ \\
Stefan-Boltzmann constant & & & & \\
\, ($\rmpi^2$/60)$k^4\!/\hbar^3c^2$ & $\sigma$ & $ 5.670\,367(13)\times 10^{-8}$ & W~m$^{-2}$~K$^{-4}$ & $ 2.3\times 10^{-6}$ \\
\multicolumn {5} {c} { \vbox to 12 pt {}
Non-SI units accepted for use with the SI} \\
electron volt ($e$/{\rm C}) {\rm J} & eV & $ 1.602\,176\,6208(98)\times 10^{-19}$ & J & $ 6.1\times 10^{-9}$ \\
(unified) atomic mass unit ${1\over12}m(^{12}$C)~~ & u & $ 1.660\,539\,040(20)\times 10^{-27}$ & kg & $ 1.2\times 10^{-8}$ \phantom{\Big|}\\
\botrule
\end{tabular}
\end{table*}

\clearpage

\def\lbar{\lambda\hskip-4.5pt\vrule height4.6pt depth-4.3pt width4pt}
\def\b{\hbox to 12pt{}}
\def\s#1{\hbox to #1pt{}}
\shortcites{2010129}
\begin{longtable*}{lllll}
\caption{The CODATA recommended values of the
fundamental constants of physics and chemistry based on the
2014 adjustment.\tlabel{tab:constants}} \\
\toprule
& & & & Relative std. \\
\s{35}Quantity & \s{-15} Symbol & \s{15} Numerical value & \s{5}Unit & uncert. $u_{\rm r}$ \\
\colrule
\endfirsthead

\caption{{\it (Continued).}} \\
\colrule
& & & & Relative std. \\
\s{35}Quantity & \s{-15} Symbol & \s{15} Numerical value & \s{5}Unit & uncert. $u_{\rm r}$ \\
\colrule
\endhead
\colrule
\endfoot
\endlastfoot
\multicolumn {5} {c} { \vbox to 12 pt {}
UNIVERSAL} \\
speed of light in vacuum & $ c,c_0 $ & $299\,792\,$458 & m~s$^{-1}$ & exact \\
magnetic constant & $\mu_0$ & 4$\rmpi\times10^{-7}$ & N~A$^{-2}$ & \\
& & $=12.566\,370\,614...\times10^{-7}$ & N~A$^{-2}$ & exact \\
electric constant 1/$\mu_0c^2$ & $\epsilon_0$ & $8.854\,187\,817...\times10^{-12}$ & F~m$^{-1}$ & exact\\
characteristic impedance of vacuum $\mu_0c$ & $Z_0$ & $376.730\,313\,461...$ & ${\rm \Omega}$ & exact\\
Newtonian constant of gravitation & $ G $ & $ 6.674\,08(31)\times 10^{-11}$ & m$^3$~kg$^{-1}$~s$^{-2}$ & $ 4.7\times 10^{-5}$ \\
& $G/\hbar c $ & $ 6.708\,61(31)\times 10^{-39}$ & $ ({\rm GeV}/c^2)^{-2}$ & $ 4.7\times 10^{-5}$ \\
Planck constant & $ h $ & $ 6.626\,070\,040(81)\times 10^{-34}$ & J~s & $ 1.2\times 10^{-8}$ \\
\b\b & & $ 4.135\,667\,662(25)\times 10^{-15}$ & eV~s & $ 6.1\times 10^{-9}$ \\
\b $h/2\rmpi$ & $\hbar$ & $ 1.054\,571\,800(13)\times 10^{-34}$ & J~s & $ 1.2\times 10^{-8}$ \\
\b\b & & $ 6.582\,119\,514(40)\times 10^{-16}$ & eV~s & $ 6.1\times 10^{-9}$ \\
\b & $\hbar c$ & $ 197.326\,9788(12)$ & MeV~fm & $ 6.1\times 10^{-9}$ \\
Planck mass~$(\hbar c/G)^{1/2}$ & $m_{\rm P}$ & $ 2.176\,470(51)\times 10^{-8}$ & kg & $ 2.3\times 10^{-5}$ \\
\b energy equivalent & $m_{\rm P}c^2$ & $ 1.220\,910(29)\times 10^{19}$ & GeV & $ 2.3\times 10^{-5}$ \\
Planck temperature~$(\hbar c^5/G)^{1/2}/k$ & $T_{\rm P}$ & $ 1.416\,808(33)\times 10^{32}$ & K & $ 2.3\times 10^{-5}$ \\
Planck length~$\hbar/m_{\rm P}c=(\hbar G/c^3)^{1/2}$ & $l_{\rm P}$ & $ 1.616\,229(38)\times 10^{-35}$ & m & $ 2.3\times 10^{-5}$ \\
Planck time $l_{\rm P}/c=(\hbar G/c^5)^{1/2}$ & $t_{\rm P}$ & $ 5.391\,16(13)\times 10^{-44}$ & s & $ 2.3\times 10^{-5}$ \\
\multicolumn {5} {c} { \vbox to 12 pt {}
ELECTROMAGNETIC} \\
elementary charge & $e$ & $ 1.602\,176\,6208(98)\times 10^{-19}$ & C & $ 6.1\times 10^{-9}$ \\
& $e/h$ & $ 2.417\,989\,262(15)\times 10^{14}$ & A~J$^{-1}$ & $ 6.1\times 10^{-9}$ \\
magnetic flux quantum $h/2e$ & ${\it \Phi}_0$ & $ 2.067\,833\,831(13)\times 10^{-15}$ & Wb & $ 6.1\times 10^{-9}$ \\
conductance quantum $2e^2\!/h$ & $G_0$ & $ 7.748\,091\,7310(18)\times 10^{-5}$ & S & $ 2.3\times 10^{-10}$ \\
\b inverse of conductance quantum & $G_0^{-1}$ & $ 12\,906.403\,7278(29)$ & ${\rm \Omega}$ & $ 2.3\times 10^{-10}$ \\
Josephson constant\footnote{See Table~\ref{tab:adopted} for the conventional value adopted 
internationally for realizing representations of the volt using the Josephson effect.}
2$e/h$ & $K_{\rm J}$ & $ 483\,597.8525(30)\times 10^{9}$ & Hz~V$^{-1}$ & $ 6.1\times 10^{-9}$ \\
von Klitzing constant\footnote{See Table~\ref{tab:adopted}
for the conventional value adopted internationally
for realizing representations of the ohm using the quantum Hall effect.}
\, $h/e^2=\mu_0c/2\alpha$ & $R_{\rm K}$ & $ 25\,812.807\,4555(59)$ & ${\rm \Omega}$ & $ 2.3\times 10^{-10}$ \\
Bohr magneton $e\hbar/2m_{\rm e}$ & $\mu_{\rm B}$ & $ 927.400\,9994(57)\times 10^{-26}$ & J~T$^{-1}$ & $ 6.2\times 10^{-9}$ \\
\b & & $ 5.788\,381\,8012(26)\times 10^{-5}$ & eV~T$^{-1}$ & $ 4.5\times 10^{-10}$ \\
\b & $\mu_{\rm B}/h$ & $ 13.996\,245\,042(86)\times 10^{9}$ & Hz~T$^{-1}$ & $ 6.2\times 10^{-9}$ \\
\b & $\mu_{\rm B}/hc$ & $ 46.686\,448\,14(29)$ & m$^{-1}~$T$^{-1}$ & $ 6.2\times 10^{-9}$ \\
\b & $\mu_{\rm B}/k$ & $ 0.671\,714\,05(39)$ & K~T$^{-1}$ & $ 5.7\times 10^{-7}$ \\
nuclear magneton $e\hbar/2m_{\rm p}$ & $\mu_{\rm N}$ & $ 5.050\,783\,699(31)\times 10^{-27}$ & J~T$^{-1}$ & $ 6.2\times 10^{-9}$ \\
\b & & $ 3.152\,451\,2550(15)\times 10^{-8}$ & eV~T$^{-1}$ & $ 4.6\times 10^{-10}$ \\
\b & $\mu_{\rm N}/h$ & $ 7.622\,593\,285(47)$ & MHz~T$^{-1}$ & $ 6.2\times 10^{-9}$ \\
\b & $\mu_{\rm N}/hc$ & $ 2.542\,623\,432(16)\times 10^{-2}$ & m$^{-1}~$T$^{-1}$ & $ 6.2\times 10^{-9}$ \\
\b & $\mu_{\rm N}/k$ & $ 3.658\,2690(21)\times 10^{-4}$ & K~T$^{-1}$ & $ 5.7\times 10^{-7}$ \\
\multicolumn {5} {c} { \vbox to 12 pt {}
ATOMIC AND NUCLEAR} \\
\multicolumn {5} {c} {General} \\
fine-structure constant $e^2\!/4\rmpi\epsilon_0\hbar c$ & $\alpha$ & $ 7.297\,352\,5664(17)\times 10^{-3}$ & & $ 2.3\times 10^{-10}$ \\
\b inverse fine-structure constant & $\alpha^{-1}$ & $ 137.035\,999\,139(31)$ & & $ 2.3\times 10^{-10}$ \\
Rydberg constant $\alpha^{2}m_{\rm e}c/2h$ & $R_\infty$ & $ 10\,973\,731.568\,508(65)$ & m$^{-1}$ & $ 5.9\times 10^{-12}$ \\
\b & $R_\infty c$ & $ 3.289\,841\,960\,355(19)\times 10^{15}$ & Hz & $ 5.9\times 10^{-12}$ \\
\b & $R_\infty hc$ & $ 2.179\,872\,325(27)\times 10^{-18}$ & J & $ 1.2\times 10^{-8}$ \\
\b & & $ 13.605\,693\,009(84)$ & eV & $ 6.1\times 10^{-9}$ \\
Bohr radius $\alpha/4\rmpi R_\infty=4\rmpi\epsilon_0\hbar^2\!/m_{\rm e}e^2$ & $a_{\rm 0}$ & $ 0.529\,177\,210\,67(12)\times 10^{-10}$ & m & $ 2.3\times 10^{-10}$ \\
Hartree energy $e^2\!/4\rmpi\epsilon_{\rm 0}a_{\rm 0}=2R_\infty hc =\alpha^2m_{\rm e}c^2$ & $E_{\rm h}$ & $ 4.359\,744\,650(54)\times 10^{-18}$ & J & $ 1.2\times 10^{-8}$ \\
\b & & $ 27.211\,386\,02(17)$ & eV & $ 6.1\times 10^{-9}$ \\
quantum of circulation & $h/2m_{\rm e}$ & $ 3.636\,947\,5486(17)\times 10^{-4}$ & m$^2~$s$^{-1}$ & $ 4.5\times 10^{-10}$ \\
& $h/m_{\rm e}$ & $ 7.273\,895\,0972(33)\times 10^{-4}$ & m$^2~$s$^{-1}$ & $ 4.5\times 10^{-10}$ \\
\multicolumn {5} {c} { \vbox to 12 pt {}
Electroweak} \\
Fermi coupling constant\footnote{Value recommended by the Particle Data
Group \numcite{2014154}.} & $G_{\rm F}/(\hbar c)^3$ & $ 1.166\,3787(6)\times 10^{-5}$ & GeV$^{-2}$ & $ 5.1\times 10^{-7}$ \vbox to 10 pt {}\\
weak mixing angle\footnote{Based on the ratio of the masses of the W and Z bosons $m_{\rm W}/m_{\rm Z}$ recommended by the Particle Data Group \numcite{2014154}.
The value for ${\rm sin}^2{\theta}_{\rm W}$
they recommend, which
is based on a particular variant of the modified minimal subtraction
$({\scriptstyle {\rm \overline{MS}}})$ scheme, is
${\rm sin}^2\hat{\theta}_{\rm W}(M_{\rm Z}) = 0.231\,26(5)$.}
$\theta_{\rm W}$ (on-shell scheme) & & & & \\
\, $\sin^2\theta_{\rm W} = s^2_{\rm W} \equiv 1-(m_{\rm W}/m_{\rm Z})^2$ & $\sin^2\theta_{\rm W}$ & $ 0.2223(21)$ & & $ 9.5\times 10^{-3}$ \\
\multicolumn {5} {c} {\vbox to 12 pt {}
Electron, e$^-$} \\
electron mass & $m_{\rm e}$ & $ 9.109\,383\,56(11)\times 10^{-31}$ & kg & $ 1.2\times 10^{-8}$ \\
\b\b & & $ 5.485\,799\,090\,70(16)\times 10^{-4}$ & u & $ 2.9\times 10^{-11}$ \\
\b energy equivalent & $m_{\rm e}c^2$ & $ 8.187\,105\,65(10)\times 10^{-14}$ & J & $ 1.2\times 10^{-8}$ \\
\b\b & & $ 0.510\,998\,9461(31)$ & MeV & $ 6.2\times 10^{-9}$ \\
electron-muon mass ratio & $m_{\rm e}/m_{\rmssmu}$ & $ 4.836\,331\,70(11)\times 10^{-3}$ & & $ 2.2\times 10^{-8}$ \\
electron-tau mass ratio & $m_{\rm e}/m_{\rmsstau}$ & $ 2.875\,92(26)\times 10^{-4}$ & & $ 9.0\times 10^{-5}$ \\
electron-proton mass ratio & $m_{\rm e}/m_{\rm p}$ & $ 5.446\,170\,213\,52(52)\times 10^{-4}$ & & $ 9.5\times 10^{-11}$ \\
electron-neutron mass ratio & $m_{\rm e}/m_{\rm n}$ & $ 5.438\,673\,4428(27)\times 10^{-4}$ & & $ 4.9\times 10^{-10}$ \\
electron-deuteron mass ratio & $m_{\rm e}/m_{\rm d}$ & $ 2.724\,437\,107\,484(96)\times 10^{-4}$ & & $ 3.5\times 10^{-11}$ \\
electron-triton mass ratio & $m_{\rm e}/m_{\rm t}$ & $ 1.819\,200\,062\,203(84)\times 10^{-4}$ & & $ 4.6\times 10^{-11}$ \\
electron-helion mass ratio & $m_{\rm e}/m_{\rm h}$ & $ 1.819\,543\,074\,854(88)\times 10^{-4}$ & & $ 4.9\times 10^{-11}$ \\
electron to alpha particle mass ratio & $m_{\rm e}/m_{\rmssalpha}$ & $ 1.370\,933\,554\,798(45)\times 10^{-4}$ & & $ 3.3\times 10^{-11}$ \\
electron charge to mass quotient & $-e/m_{\rm e}$ & $ -1.758\,820\,024(11)\times 10^{11}$ & C~kg$^{-1}$ & $ 6.2\times 10^{-9}$ \\
electron molar mass $N_{\rm A}m_{\rm e}$& $M({\rm e}),M_{\rm e}$ & $ 5.485\,799\,090\,70(16)\times 10^{-7}$ & kg mol$^{-1}$ & $ 2.9\times 10^{-11}$ \\
Compton wavelength $h/m_{\rm e}c$ & $\lambda_{\rm C}$ & $ 2.426\,310\,2367(11)\times 10^{-12}$ & m & $ 4.5\times 10^{-10}$ \\
\b $\lambda_{\rm C}/2\rmpi=\alpha a_{\rm 0}=\alpha^2\!/4\rmpi R_\infty$ & $\lbar_{\rm C}$ & $ 386.159\,267\,64(18)\times 10^{-15}$ & m & $ 4.5\times 10^{-10}$ \\
 classical electron radius $\alpha^2a_{\rm 0}$ & $r_{\rm e}$ & $ 2.817\,940\,3227(19)\times 10^{-15}$ & m & $ 6.8\times 10^{-10}$ \\
Thomson cross section (8$\rmpi/3)r^2_{\rm e}$ & $\sigma_{\rm e}$ & $ 0.665\,245\,871\,58(91)\times 10^{-28}$ & m$^2$ & $ 1.4\times 10^{-9}$ \\
electron magnetic moment & $\mu_{\rm e}$ & $ -928.476\,4620(57)\times 10^{-26}$ & J~T$^{-1}$ & $ 6.2\times 10^{-9}$ \\
\b to Bohr magneton ratio & $\mu_{\rm e}/\mu_{\rm B}$ & $ -1.001\,159\,652\,180\,91(26)$ & & $ 2.6\times 10^{-13}$ \\
\b to nuclear magneton ratio & $\mu_{\rm e}/\mu_{\rm N}$ & $ -1838.281\,972\,34(17)$ & & $ 9.5\times 10^{-11}$ \\
electron magnetic moment & & & & \\
\, anomaly $|\mu_{\rm e}|/\mu_{\rm B}-1$ & $a_{\rm e}$ & $ 1.159\,652\,180\,91(26)\times 10^{-3}$ & & $ 2.3\times 10^{-10}$ \\
electron $g$-factor $-2(1+a_{\rm e})$ & $g_{\rm e}$ & $ -2.002\,319\,304\,361\,82(52)$ & & $ 2.6\times 10^{-13}$ \\
electron-muon magnetic moment ratio & $\mu_{\rm e}/\mu_{\rmssmu}$ & $ 206.766\,9880(46)$ & & $ 2.2\times 10^{-8}$ \\
electron-proton magnetic moment ratio & $\mu_{\rm e}/\mu_{\rm p}$ & $ -658.210\,6866(20)$ & & $ 3.0\times 10^{-9}$ \\
electron to shielded proton magnetic& & & & \\
\, moment ratio (H$_2$O, sphere, 25 $^\circ$C) & $\mu_{\rm e}/\mu^\prime_{\rm p}$ & $ -658.227\,5971(72)$ & & $ 1.1\times 10^{-8}$ \\
electron-neutron magnetic moment ratio & $\mu_{\rm e}/\mu_{\rm n}$ & $ 960.920\,50(23)$ & & $ 2.4\times 10^{-7}$ \\
electron-deuteron magnetic moment ratio & $\mu_{\rm e}/\mu_{\rm d}$ & $ -2143.923\,499(12)$ & & $ 5.5\times 10^{-9}$ \\
electron to shielded helion magnetic & & & & \\
\, moment ratio (gas, sphere, 25 $^\circ$C) & $\mu_{\rm e}/\mu^\prime_{\rm h}$ & $ 864.058\,257(10)$ & & $ 1.2\times 10^{-8}$ \\
electron gyromagnetic ratio $2|\mu_{\rm e}|/\hbar$ & $\gamma_{\rm e}$ & $ 1.760\,859\,644(11)\times 10^{11}$ & s$^{-1}~$T$^{-1}$ & $ 6.2\times 10^{-9}$ \\
& $\gamma_{\rm e}/2\rmpi$ & $ 28\,024.951\,64(17)$ & MHz~T$^{-1}$ & $ 6.2\times 10^{-9}$ \\
\multicolumn {5} {c} { \vbox to 12 pt {}
Muon, ${\rmmu}^-$} \\
muon mass & $m_{\rmssmu}$ & $ 1.883\,531\,594(48)\times 10^{-28}$ & kg & $ 2.5\times 10^{-8}$ \\
& & $ 0.113\,428\,9257(25)$ & u & $ 2.2\times 10^{-8}$ \\
\b energy equivalent & $m_{\rmssmu}c^2$ & $ 1.692\,833\,774(43)\times 10^{-11}$ & J & $ 2.5\times 10^{-8}$ \\
& & $ 105.658\,3745(24)$ & MeV & $ 2.3\times 10^{-8}$ \\
muon-electron mass ratio & $m_{\rmssmu}/m_{\rm e}$ & $ 206.768\,2826(46)$ & & $ 2.2\times 10^{-8}$ \\
muon-tau mass ratio & $m_{\rmssmu}/m_{\rmsstau}$ & $ 5.946\,49(54)\times 10^{-2}$ & & $ 9.0\times 10^{-5}$ \\
muon-proton mass ratio & $m_{\rmssmu}/m_{\rm p}$ & $ 0.112\,609\,5262(25)$ & & $ 2.2\times 10^{-8}$ \\
muon-neutron mass ratio & $m_{\rmssmu}/m_{\rm n}$ & $ 0.112\,454\,5167(25)$ & & $ 2.2\times 10^{-8}$ \\
muon molar mass $N_{\rm A}m_{\rmssmu}$& $M({\rmmu}),M_{\rmssmu}$ & $ 0.113\,428\,9257(25)\times 10^{-3}$ & kg mol$^{-1}$ & $ 2.2\times 10^{-8}$ \\
muon Compton wavelength $h/m_{\rmssmu}c$ & $\lambda_{{\rm C},{\rmssmu}}$ & $ 11.734\,441\,11(26)\times 10^{-15}$ & m & $ 2.2\times 10^{-8}$ \\
\b $\lambda_{{\rm C},{\rmssmu}}/2\rmpi$ & $\lbar_{{\rm C},{\rmssmu}}$ & $ 1.867\,594\,308(42)\times 10^{-15}$ & m & $ 2.2\times 10^{-8}$ \\
muon magnetic moment & $\mu_{\rmssmu}$ & $ -4.490\,448\,26(10)\times 10^{-26}$ & J~T$^{-1}$ & $ 2.3\times 10^{-8}$ \\
\b to Bohr magneton ratio & $\mu_{\rmssmu}/\mu_{\rm B}$ & $ -4.841\,970\,48(11)\times 10^{-3}$ & & $ 2.2\times 10^{-8}$ \\
\b to nuclear magneton ratio & $\mu_{\rmssmu}/\mu_{\rm N}$ & $ -8.890\,597\,05(20)$ & & $ 2.2\times 10^{-8}$ \\
muon magnetic moment anomaly & & & & \\
\, $|\mu_{\rmssmu}|/(e\hbar/2m_{\rmssmu})-1$ & $a_{\rmssmu}$ & $ 1.165\,920\,89(63)\times 10^{-3}$ & & $ 5.4\times 10^{-7}$ \\
muon $g$-factor $-2(1+a_{\rmssmu}$) & $g_{\rmssmu}$ & $ -2.002\,331\,8418(13)$ & & $ 6.3\times 10^{-10}$ \\
muon-proton magnetic moment ratio & $\mu_{\rmssmu}/\mu_{\rm p}$ & $ -3.183\,345\,142(71)$ & & $ 2.2\times 10^{-8}$ \\
\multicolumn {5} {c} { \vbox to 12 pt {}
Tau, ${\rmtau}^-$} \\
tau mass\footnote{This and all other values involving $m_{\rmsstau}$ are based on
the value of $m_{\rmsstau}c^2$ in MeV recommended by the
Particle Data Group \numcite{2014154}.}
& $m_{\rmsstau}$ & $ 3.167\,47(29)\times 10^{-27}$ & kg & $ 9.0\times 10^{-5}$ \\
& & $ 1.907\,49(17)$ & u & $ 9.0\times 10^{-5}$ \\
\b energy equivalent & $m_{\rmsstau}c^2$ & $ 2.846\,78(26)\times 10^{-10}$ & J & $ 9.0\times 10^{-5}$ \\
& & $ 1776.82(16)$ & MeV & $ 9.0\times 10^{-5}$ \\
tau-electron mass ratio & $m_{\rmsstau}/m_{\rm e}$ & $ 3477.15(31)$ & & $ 9.0\times 10^{-5}$ \\
tau-muon mass ratio & $m_{\rmsstau}/m_{\rmssmu}$ & $ 16.8167(15)$ & & $ 9.0\times 10^{-5}$ \\
tau-proton mass ratio & $m_{\rmsstau}/m_{\rm p}$ & $ 1.893\,72(17)$ & & $ 9.0\times 10^{-5}$ \\
tau-neutron mass ratio & $m_{\rmsstau}/m_{\rm n}$ & $ 1.891\,11(17)$ & & $ 9.0\times 10^{-5}$ \\
tau molar mass $N_{\rm A}m_{\rmsstau}$& $M({\rmtau}),M_{\rmsstau}$ & $ 1.907\,49(17)\times 10^{-3}$ & kg mol$^{-1}$ & $ 9.0\times 10^{-5}$ \\
tau Compton wavelength $h/m_{\rmsstau}c$ & $\lambda_{{\rm C},{\rmsstau}}$ & $ 0.697\,787(63)\times 10^{-15}$ & m & $ 9.0\times 10^{-5}$ \\
\b $\lambda_{{\rm C},{\rmsstau}}/2\rmpi$ & $\lbar_{{\rm C},{\rmsstau}}$ & $ 0.111\,056(10)\times 10^{-15}$ & m & $ 9.0\times 10^{-5}$ \\
\multicolumn {5} {c} {\vbox to 12 pt {} Proton, p} \\
proton mass & $m_{\rm p}$ & $ 1.672\,621\,898(21)\times 10^{-27}$ & kg & $ 1.2\times 10^{-8}$ \\
& & $ 1.007\,276\,466\,879(91)$ & u & $ 9.0\times 10^{-11}$ \\
\b energy equivalent & $m_{\rm p}c^2$ & $ 1.503\,277\,593(18)\times 10^{-10}$ & J & $ 1.2\times 10^{-8}$ \\
& & $ 938.272\,0813(58)$ & MeV & $ 6.2\times 10^{-9}$ \\
proton-electron mass ratio & $m_{\rm p}/m_{\rm e}$ & $ 1836.152\,673\,89(17)$ & & $ 9.5\times 10^{-11}$ \\
proton-muon mass ratio & $m_{\rm p}/m_{\rmssmu}$ & $ 8.880\,243\,38(20)$ & & $ 2.2\times 10^{-8}$ \\
proton-tau mass ratio & $m_{\rm p}/m_{\rmsstau}$ & $ 0.528\,063(48)$ & & $ 9.0\times 10^{-5}$ \\
proton-neutron mass ratio & $m_{\rm p}/m_{\rm n}$ & $ 0.998\,623\,478\,44(51)$ & & $ 5.1\times 10^{-10}$ \\
proton charge to mass quotient & $e/m_{\rm p}$ & $ 9.578\,833\,226(59)\times 10^{7}$ & C kg$^{-1}$ & $ 6.2\times 10^{-9}$ \\
proton molar mass $N_{\rm A}m_{\rm p}$& $M$(p), $M_{\rm p}$ & $ 1.007\,276\,466\,879(91)\times 10^{-3}$ & kg mol$^{-1}$ & $ 9.0\times 10^{-11}$ \\
proton Compton wavelength $h/m_{\rm p}c$ & $\lambda_{\rm C,p}$ & $ 1.321\,409\,853\,96(61)\times 10^{-15}$ & m & $ 4.6\times 10^{-10}$ \\
\b $\lambda_{\rm C,p}/2\rmpi$ & $\lbar_{\rm C,p}$ & $ 0.210\,308\,910\,109(97)\times 10^{-15}$ & m & $ 4.6\times 10^{-10}$ \\
proton rms charge radius & $r_{\rm p}$ & $ 0.8751(61)\times 10^{-15}$ & m & $ 7.0\times 10^{-3}$ \\
proton magnetic moment & $\mu_{\rm p}$ & $ 1.410\,606\,7873(97)\times 10^{-26}$ & J~T$^{-1}$ & $ 6.9\times 10^{-9}$ \\
\b to Bohr magneton ratio & $\mu_{\rm p}/\mu_{\rm B}$ & $ 1.521\,032\,2053(46)\times 10^{-3}$ & & $ 3.0\times 10^{-9}$ \\
\b to nuclear magneton ratio & $\mu_{\rm p}/\mu_{\rm N}$ & $ 2.792\,847\,3508(85)$ & & $ 3.0\times 10^{-9}$ \\
proton $g$-factor $2\mu_{\rm p}/\mu_{\rm N}$ & $g_{\rm p}$ & $ 5.585\,694\,702(17)$ & & $ 3.0\times 10^{-9}$ \\
proton-neutron magnetic moment ratio & $\mu_{\rm p}/\mu_{\rm n}$ & $ -1.459\,898\,05(34)$ & & $ 2.4\times 10^{-7}$ \\
shielded proton magnetic moment & $\mu^\prime_{\rm p}$ & $ 1.410\,570\,547(18)\times 10^{-26}$ & J~T$^{-1}$ & $ 1.3\times 10^{-8}$ \\
\, (H$_{2}$O, sphere, 25 $^\circ$C) & & & & \\
\b to Bohr magneton ratio & $\mu^\prime_{\rm p}/\mu_{\rm B}$ & $ 1.520\,993\,128(17)\times 10^{-3}$ & & $ 1.1\times 10^{-8}$ \\
\b to nuclear magneton ratio & $\mu^\prime_{\rm p}/\mu_{\rm N}$ & $ 2.792\,775\,600(30)$ & & $ 1.1\times 10^{-8}$ \\
proton magnetic shielding correction& & & & \\
\, $1-\mu^\prime_{\rm p}/\mu_{\rm p}$ \ (H$_{2}$O, sphere, 25 $^\circ$C) & $\sigma^\prime_{\rm p}$ & $ 25.691(11)\times 10^{-6}$ & &  $ 4.4\times 10^{-4}$ \\
proton gyromagnetic ratio $2\mu_{\rm p}/\hbar$ & $\gamma_{\rm p}$ & $ 2.675\,221\,900(18)\times 10^{8}$ & s$^{-1}~$T$^{-1}$ & $ 6.9\times 10^{-9}$ \\
& $\gamma_{\rm p}/2\rmpi$ & $ 42.577\,478\,92(29)$ & MHz~T$^{-1}$ & $ 6.9\times 10^{-9}$ \\
shielded proton gyromagnetic ratio& & & & \\
$2\mu^\prime_{\rm p}/\hbar$ \ (H$_{2}$O, sphere, 25 $^\circ$C)& $\gamma^\prime_{\rm p}$ & $ 2.675\,153\,171(33)\times 10^{8}$ & s$^{-1}~$T$^{-1}$ & $ 1.3\times 10^{-8}$ \\
& $\gamma^\prime_{\rm p}/2\rmpi$ & $ 42.576\,385\,07(53)$ & MHz~T$^{-1}$ & $ 1.3\times 10^{-8}$ \\
\multicolumn {5} {c} {\vbox to 12 pt {} Neutron, n} \\
neutron mass & $m_{\rm n}$ & $ 1.674\,927\,471(21)\times 10^{-27}$ & kg & $ 1.2\times 10^{-8}$ \\
& & $ 1.008\,664\,915\,88(49)$ & u & $ 4.9\times 10^{-10}$ \\
\b energy equivalent & $m_{\rm n}c^2$ & $ 1.505\,349\,739(19)\times 10^{-10}$ & J & $ 1.2\times 10^{-8}$ \\
& & $ 939.565\,4133(58)$ & MeV & $ 6.2\times 10^{-9}$ \\
neutron-electron mass ratio & $m_{\rm n}/m_{\rm e}$ & $ 1838.683\,661\,58(90)$ & & $ 4.9\times 10^{-10}$ \\
neutron-muon mass ratio & $m_{\rm n}/m_{\rmssmu}$ & $ 8.892\,484\,08(20)$ & & $ 2.2\times 10^{-8}$ \\
neutron-tau mass ratio & $m_{\rm n}/m_{\rmsstau}$ & $ 0.528\,790(48)$ & & $ 9.0\times 10^{-5}$ \\
neutron-proton mass ratio & $m_{\rm n}/m_{\rm p}$ & $ 1.001\,378\,418\,98(51)$ & & $ 5.1\times 10^{-10}$ \\
neutron-proton mass difference & $m_{\rm n}-m_{\rm p}$ & $ 2.305\,573\,77(85)\times 10^{-30}$ & kg & $ 3.7\times 10^{-7}$  \\
& & $ 0.001\,388\,449\,00(51)$ & u & $ 3.7\times 10^{-7}$ \\
\b energy equivalent & ($m_{\rm n}-m_{\rm p})c^2$~~ & $ 2.072\,146\,37(76)\times 10^{-13}$ & J & $ 3.7\times 10^{-7}$ \\
& & $ 1.293\,332\,05(48)$ & MeV & $ 3.7\times 10^{-7}$ \\
neutron molar mass $N_{\rm A}m_{\rm n}$ & $M({\rm n}),M_{\rm n}$ & $ 1.008\,664\,915\,88(49)\times 10^{-3}$ & kg mol$^{-1}$ & $ 4.9\times 10^{-10}$ \\
neutron Compton wavelength $h/m_{\rm n}c$ & $\lambda_{\rm C,n}$ & $ 1.319\,590\,904\,81(88)\times 10^{-15}$ & m & $ 6.7\times 10^{-10}$ \\
\b $\lambda_{\rm C,n}/2\rmpi$ & $\lbar_{\rm C,n}$ & $ 0.210\,019\,415\,36(14)\times 10^{-15}$ & m & $ 6.7\times 10^{-10}$ \\
neutron magnetic moment & $\mu_{\rm n}$ & $ -0.966\,236\,50(23)\times 10^{-26}$ & J~T$^{-1}$ & $ 2.4\times 10^{-7}$ \\
\b to Bohr magneton ratio & $\mu_{\rm n}/\mu_{\rm B}$ & $ -1.041\,875\,63(25)\times 10^{-3}$ & & $ 2.4\times 10^{-7}$ \\
\b to nuclear magneton ratio & $\mu_{\rm n}/\mu_{\rm N}$ & $ -1.913\,042\,73(45)$ & & $ 2.4\times 10^{-7}$ \\
neutron $g$-factor $2\mu_{\rm n}/\mu_{\rm N}$ & $g_{\rm n}$ & $ -3.826\,085\,45(90)$ & & $ 2.4\times 10^{-7}$ \\
neutron-electron magnetic moment ratio & $\mu_{\rm n}/\mu_{\rm e}$ & $ 1.040\,668\,82(25)\times 10^{-3}$ & & $ 2.4\times 10^{-7}$ \\
neutron-proton magnetic moment ratio & $\mu_{\rm n}/\mu_{\rm p}$ & $ -0.684\,979\,34(16)$ & & $ 2.4\times 10^{-7}$ \\
neutron to shielded proton magnetic& & & & \\
\, moment ratio \ (H$_2$O, sphere, 25 $^\circ$C)& $\mu_{\rm n}/\mu_{\rm p}^\prime$ & $ -0.684\,996\,94(16)$ & & $ 2.4\times 10^{-7}$ \\
neutron gyromagnetic ratio $2|\mu_{\rm n}|/\hbar$ & $\gamma_{\rm n}$ & $ 1.832\,471\,72(43)\times 10^{8}$ & s$^{-1}~$T$^{-1}$ & $ 2.4\times 10^{-7}$ \\
& $\gamma_{\rm n}/2\rmpi$ & $ 29.164\,6933(69)$ & MHz~T$^{-1}$ & $ 2.4\times 10^{-7}$ \\
\multicolumn {5} {c} {\vbox to 12 pt {} Deuteron, d} \\
deuteron mass & $m_{\rm d}$ & $ 3.343\,583\,719(41)\times 10^{-27}$ & kg & $ 1.2\times 10^{-8}$ \\
& & $ 2.013\,553\,212\,745(40)$ & u & $ 2.0\times 10^{-11}$ \\
\b energy equivalent & $m_{\rm d}c^2$ & $ 3.005\,063\,183(37)\times 10^{-10}$ & J & $ 1.2\times 10^{-8}$ \\
& & $ 1875.612\,928(12)$ & MeV & $ 6.2\times 10^{-9}$ \\
deuteron-electron mass ratio & $m_{\rm d}/m_{\rm e}$ & $ 3670.482\,967\,85(13)$ & & $ 3.5\times 10^{-11}$ \\
deuteron-proton mass ratio & $m_{\rm d}/m_{\rm p}$ & $ 1.999\,007\,500\,87(19)$ & & $ 9.3\times 10^{-11}$ \\
deuteron molar mass $N_{\rm A}m_{\rm d}$& $M({\rm d}),M_{\rm d}$ & $ 2.013\,553\,212\,745(40)\times 10^{-3}$ & kg mol$^{-1}$ & $ 2.0\times 10^{-11}$ \\
deuteron rms charge radius & $r_{\rm d}$ & $ 2.1413(25)\times 10^{-15}$ & m & $ 1.2\times 10^{-3}$ \\
deuteron magnetic moment & $\mu_{\rm d}$ & $ 0.433\,073\,5040(36)\times 10^{-26}$ & J~T$^{-1}$ & $ 8.3\times 10^{-9}$ \\
\b to Bohr magneton ratio & $\mu_{\rm d}/\mu_{\rm B}$ & $ 0.466\,975\,4554(26)\times 10^{-3}$ & & $ 5.5\times 10^{-9}$ \\
\b to nuclear magneton ratio & $\mu_{\rm d}/\mu_{\rm N}$ & $ 0.857\,438\,2311(48)$ & & $ 5.5\times 10^{-9}$ \\
deuteron $g$-factor $\mu_{\rm d}/\mu_{\rm N}$ & $g_{\rm d}$ & $ 0.857\,438\,2311(48)$ & & $ 5.5\times 10^{-9}$ \\
deuteron-electron magnetic moment ratio & $\mu_{\rm d}/\mu_{\rm e}$ & $ -4.664\,345\,535(26)\times 10^{-4}$ & & $ 5.5\times 10^{-9}$ \\
deuteron-proton magnetic moment ratio & $\mu_{\rm d}/\mu_{\rm p}$ & $ 0.307\,012\,2077(15)$ & & $ 5.0\times 10^{-9}$ \\
deuteron-neutron magnetic moment ratio & $\mu_{\rm d}/\mu_{\rm n}$ & $ -0.448\,206\,52(11)$ & & $ 2.4\times 10^{-7}$ \\
\multicolumn {5} {c} {\vbox to 12 pt {} Triton, t} \\
triton mass & $m_{\rm t}$ & $ 5.007\,356\,665(62)\times 10^{-27}$ & kg & $ 1.2\times 10^{-8}$ \\
& & $ 3.015\,500\,716\,32(11)$ & u & $ 3.6\times 10^{-11}$ \\
\b energy equivalent & $m_{\rm t}c^2$ & $ 4.500\,387\,735(55)\times 10^{-10}$ & J & $ 1.2\times 10^{-8}$ \\
& & $ 2808.921\,112(17)$ & MeV & $ 6.2\times 10^{-9}$ \\
triton-electron mass ratio & $m_{\rm t}/m_{\rm e}$ & $ 5496.921\,535\,88(26)$ & & $ 4.6\times 10^{-11}$ \\
triton-proton mass ratio & $m_{\rm t}/m_{\rm p}$ & $ 2.993\,717\,033\,48(22)$ & & $ 7.5\times 10^{-11}$ \\
triton molar mass $N_{\rm A}m_{\rm t}$& $M({\rm t}),M_{\rm t}$ & $ 3.015\,500\,716\,32(11)\times 10^{-3}$ & kg mol$^{-1}$ & $ 3.6\times 10^{-11}$ \\
triton magnetic moment & $\mu_{\rm t}$ & $ 1.504\,609\,503(12)\times 10^{-26}$ & J~T$^{-1}$ & $ 7.8\times 10^{-9}$ \\
\b to Bohr magneton ratio & $\mu_{\rm t}/\mu_{\rm B}$ & $ 1.622\,393\,6616(76)\times 10^{-3}$ & & $ 4.7\times 10^{-9}$ \\
\b to nuclear magneton ratio & $\mu_{\rm t}/\mu_{\rm N}$ & $ 2.978\,962\,460(14)$ & & $ 4.7\times 10^{-9}$ \\
triton $g$-factor $2\mu_{\rm t}/\mu_{\rm N}$ & $g_{\rm t}$ & $ 5.957\,924\,920(28)$ & & $ 4.7\times 10^{-9}$ \\
\multicolumn {5} {c} {\vbox to 12 pt {} Helion, h} \\
helion mass & $m_{\rm h}$ & $ 5.006\,412\,700(62)\times 10^{-27}$ & kg & $ 1.2\times 10^{-8}$ \\
& & $ 3.014\,932\,246\,73(12)$ & u & $ 3.9\times 10^{-11}$ \\
\b energy equivalent & $m_{\rm h}c^2$ & $ 4.499\,539\,341(55)\times 10^{-10}$ & J & $ 1.2\times 10^{-8}$ \\
& & $ 2808.391\,586(17)$ & MeV & $ 6.2\times 10^{-9}$ \\
helion-electron mass ratio & $m_{\rm h}/m_{\rm e}$ & $ 5495.885\,279\,22(27)$ & & $ 4.9\times 10^{-11}$ \\
helion-proton mass ratio & $m_{\rm h}/m_{\rm p}$ & $ 2.993\,152\,670\,46(29)$ & & $ 9.6\times 10^{-11}$ \\
helion molar mass $N_{\rm A}m_{\rm h}$& $M({\rm h}),M_{\rm h}$ & $ 3.014\,932\,246\,73(12)\times 10^{-3}$ & kg mol$^{-1}$ & $ 3.9\times 10^{-11}$ \\
helion magnetic moment & $\mu_{\rm h}$ & $ -1.074\,617\,522(14)\times 10^{-26}$ & J~T$^{-1}$ & $ 1.3\times 10^{-8}$ \\
\b to Bohr magneton ratio & $\mu_{\rm h}/\mu_{\rm B}$ & $ -1.158\,740\,958(14)\times 10^{-3}$ & & $ 1.2\times 10^{-8}$ \\
\b to nuclear magneton ratio & $\mu_{\rm h}/\mu_{\rm N}$ & $ -2.127\,625\,308(25)$ & & $ 1.2\times 10^{-8}$ \\
helion $g$-factor $2\mu_{\rm h}/\mu_{\rm N}$ & $g_{\rm h}$ & $ -4.255\,250\,616(50)$ & & $ 1.2\times 10^{-8}$ \\
shielded helion magnetic moment & $\mu^\prime_{\rm h}$ & $ -1.074\,553\,080(14)\times 10^{-26}$ & J~T$^{-1}$ & $ 1.3\times 10^{-8}$ \\
\, (gas, sphere, 25 $^\circ$C) & & & & \\
\b to Bohr magneton ratio & $\mu^\prime_{\rm h}/\mu_{\rm B}$ & $ -1.158\,671\,471(14)\times 10^{-3}$ & & $ 1.2\times 10^{-8}$ \\
\b to nuclear magneton ratio & $\mu^\prime_{\rm h}/\mu_{\rm N}$ & $ -2.127\,497\,720(25)$ & & $ 1.2\times 10^{-8}$ \\
shielded helion to proton magnetic & & & & \\
\b moment ratio \ (gas, sphere, 25 $^\circ$C) & $\mu^\prime_{\rm h}/\mu_{\rm p}$ & $ -0.761\,766\,5603(92)$ & & $ 1.2\times 10^{-8}$ \\
shielded helion to shielded proton magnetic & & & & \\
\b moment ratio \ (gas/H$_2$O, spheres, 25 $^\circ$C) & $\mu^\prime_{\rm h}/\mu^\prime_{\rm p}$ & $ -0.761\,786\,1313(33)$ & & $ 4.3\times 10^{-9}$ \\
shielded helion gyromagnetic ratio& & & & \\
\, $2|\mu^\prime_{\rm h}|/\hbar$ \ (gas, sphere, 25 $^\circ$C) & $\gamma^\prime_{\rm h}$ & $ 2.037\,894\,585(27)\times 10^{8}$ & s$^{-1}$ T$^{-1}$ & $ 1.3\times 10^{-8}$ \\ 
& $\gamma^\prime_{\rm h}/2\rmpi$ & $ 32.434\,099\,66(43)$ & MHz~T$^{-1}$ & $ 1.3\times 10^{-8}$ \\
\multicolumn {5} {c} {\vbox to 12 pt {} Alpha particle, ${\rmalpha}$} \\
alpha particle mass & $m_{\rmssalpha}$ & $ 6.644\,657\,230(82)\times 10^{-27}$ & kg & $ 1.2\times 10^{-8}$ \\
& & $ 4.001\,506\,179\,127(63)$ & u & $ 1.6\times 10^{-11}$ \\
\b energy equivalent & $m_{\rmssalpha}c^2$ & $ 5.971\,920\,097(73)\times 10^{-10}$ & J & $ 1.2\times 10^{-8}$ \\
& & $ 3727.379\,378(23)$ & MeV & $ 6.2\times 10^{-9}$ \\
alpha particle to electron mass ratio & $m_{\rmssalpha}/m_{\rm e}$ & $ 7294.299\,541\,36(24)$ & & $ 3.3\times 10^{-11}$ \\
alpha particle to proton mass ratio & $m_{\rmssalpha}/m_{\rm p}$ & $ 3.972\,599\,689\,07(36)$ & & $ 9.2\times 10^{-11}$ \\
alpha particle molar mass $N_{\rm A}m_{\rmssalpha}$& $M({\rmalpha}),M_{\rmssalpha}$ & $ 4.001\,506\,179\,127(63)\times 10^{-3}$ & kg mol$^{-1}$ & $ 1.6\times 10^{-11}$ \\
\multicolumn {5} {c} {\vbox to 12 pt {} PHYSICOCHEMICAL} \\
Avogadro constant & $N_{\rm A},L$ & $ 6.022\,140\,857(74)\times 10^{23}$ & mol$^{-1}$ & $ 1.2\times 10^{-8}$ \\
atomic mass constant & & & & \\
\, $m_{\rm u}=\frac {1}{12}m(^{12}{\rm C})= 1$ u & $m_{\rm u}$ & $ 1.660\,539\,040(20)\times 10^{-27}$ & kg & $ 1.2\times 10^{-8}$ \\
\b energy equivalent & $m_{\rm u}c^2$ & $ 1.492\,418\,062(18)\times 10^{-10}$ & J & $ 1.2\times 10^{-8}$ \\
& & $ 931.494\,0954(57)$ & MeV & $ 6.2\times 10^{-9}$ \\
Faraday constant\footnote{The numerical value of $F$ to be used in coulometric chemical measurements
is $ 96\,485.3251(12)$~[$ 1.2\times 10^{-8}$] when the relevant current is measured
in terms of representations of the volt and ohm based on the Josephson
and quantum Hall effects and the internationally adopted conventional
values of the Josephson and von Klitzing constants $K_{\rm J-90}$ and
$R_{\rm K-90}$ given in Table~\ref{tab:adopted}.
}
$N_{\rm A}e$ & $F$ & $ 96\,485.332\,89(59)$ & C~mol$^{-1}$ & $ 6.2\times 10^{-9}$ \\
molar Planck constant & $N_{\rm A}h$ & $ 3.990\,312\,7110(18)\times 10^{-10}$ & J~s~mol$^{-1}$ & $ 4.5\times 10^{-10}$ \\
& $N_{\rm A}hc$ & $ 0.119\,626\,565\,582(54)$ & J~m~mol$^{-1}$ & $ 4.5\times 10^{-10}$ \\
molar gas constant & $R$ & $ 8.314\,4598(48)$ & J~mol$^{-1}~$K$^{-1}$ & $ 5.7\times 10^{-7}$ \\
Boltzmann constant $R/N_{\rm A}$ & $k$ & $ 1.380\,648\,52(79)\times 10^{-23}$ & J~K$^{-1}$ & $ 5.7\times 10^{-7}$ \\
& & $ 8.617\,3303(50)\times 10^{-5}$ & eV~K$^{-1}$ & $ 5.7\times 10^{-7}$ \\
\b & $k/h$ & $ 2.083\,6612(12)\times 10^{10}$ & Hz~K$^{-1}$ & $ 5.7\times 10^{-7}$ \\
\b & $k/hc$ & $ 69.503\,457(40)$ & m$^{-1}~$K$^{-1}$ & $ 5.7\times 10^{-7}$ \\
molar volume of ideal gas $RT/p$ & & & & \\
\b\b $T=273.15\ {\rm K},\,p=100\ {\rm kPa}$ & $V_{\rm m}$ & $ 22.710\,947(13)\times 10^{-3}$ & m$^{3}~$mol$^{-1}$ & $ 5.7\times 10^{-7}$ \\
\b Loschmidt constant $N_{\rm A}/V_{\rm m}$ & $n_0$ & $ 2.651\,6467(15)\times 10^{25}$ & m$^{-3}$ & $ 5.7\times 10^{-7}$ \\
molar volume of ideal gas $RT/p$ & & & & \\
\b\b $T=273.15\ {\rm K},\,p=101.325\ {\rm kPa}$ & $V_{\rm m}$ & $ 22.413\,962(13)\times 10^{-3}$ & m$^{3}~$mol$^{-1}$ & $ 5.7\times 10^{-7}$ \\
\b Loschmidt constant $N_{\rm A}/V_{\rm m}$ & $n_0$ & $ 2.686\,7811(15)\times 10^{25}$ & m$^{-3}$ & $ 5.7\times 10^{-7}$ \\
Sackur-Tetrode (absolute entropy) constant\footnote{The entropy of an ideal monoatomic gas of relative atomic mass
$A_{\rm r}$ is given by $S = S_0 +{3\over2} R\, \ln A_{\rm r}
-R\, \ln(p/p_0) +{5\over2}R\,\ln(T/{\rm K}).$
}
& & & & \\
\, $\frac {5} {2}+\ln[(2\rmpi m_{\rm u}kT_1/h^2)^{3/2}kT_1/p_0]$ & & & & \\
\b $T_1=1\ {\rm K},\,p_0\,=\,100\ {\rm kPa}$ & $S_0/R$ & $ -1.151\,7084(14)$ & & $ 1.2\times 10^{-6}$ \\
\b $T_1=1\ {\rm K},\,p_0\,=\,101.325\ {\rm kPa}$ & & $ -1.164\,8714(14)$ & & $ 1.2\times 10^{-6}$ \\
Stefan-Boltzmann constant & & & & \\
\, ($\rmpi^2/60)k^4\!/\hbar^3c^2$ & $\sigma$ & $ 5.670\,367(13)\times 10^{-8}$ & W~m$^{-2}~$K$^{-4}$ & $ 2.3\times 10^{-6}$ \\
first radiation constant 2$\rmpi hc^2$ & $c_1$ & $ 3.741\,771\,790(46)\times 10^{-16}$ & W~m$^{2}$ & $ 1.2\times 10^{-8}$ \\
first radiation constant for spectral radiance 2$hc^2$ & $c_{\rm 1L}$ & $ 1.191\,042\,953(15)\times 10^{-16}$ & W~m$^{2}$~sr$^{-1}$ & $ 1.2\times 10^{-8}$ \vbox to 10 pt{}\\
second radiation constant $hc/k$ & $c_2$ & $ 1.438\,777\,36(83)\times 10^{-2}$ & m~K & $ 5.7\times 10^{-7}$ \\
Wien displacement law constants & & & & \\
\, $b=\lambda_{\rm max}T=c_2/4.965\,114\,231...$ & $b$ & $ 2.897\,7729(17)\times 10^{-3}$ & m~K & $ 5.7\times 10^{-7}$ \\
\, $b^\prime=\nu_{\rm max}/T=2.821\,439\,372...\,c/c_2 $ & $b^\prime$ & $ 5.878\,9238(34)\times 10^{10}$ & Hz~K$^{-1}$ & $ 5.7\times 10^{-7}$ \\
\botrule
\end{longtable*}

\begin{table*}
\caption{The variances, covariances, and correlation coefficients of the values of a 
selected group of constants based on the 2014 CODATA adjustment.  The numbers in bold
above the main diagonal are $10^{16}$ times the numerical values of the relative covariances;
the numbers in bold on the main diagonal are $10^{16}$ times the numerical values of the
relative variances; and the numbers in italics below the main diagonal are the correlation
coefficients.$^1$}
\tlabel{tab:varmatrix}
\def\sp{\hbox to 30.7 pt{}}
\hbox to 6.8 in{
\begin{tabular} {c@{\sp}r@{\sp}r@{\sp}r@{\sp}r@{\sp}r@{\sp}r@{\sp}r}
\toprule
& $\alpha$~~~ & $h$~~~  & $e$~~~  & $m_{\rm e}$~~~ & $N_{\rm A}$~~~  & $m_{\rm e}/m_{\rm \mu}$  & $F$~~~ \\
\colrule
$\alpha$ & ${\bf  0.0005}$ & ${\bf  0.0005}$ & ${\bf
 0.0005}$ & ${\bf  -0.0005}$ & ${\bf  0.0005}$ &
${\bf  -0.0010}$ & ${\bf  0.0010}$ \vbox to 10 pt {}\\
$h$ & ${\it  0.0176}$ & ${\bf  1.5096}$ & ${\bf  0.7550}$ & ${\bf  1.5086}$ & ${\bf  -1.5086}$ & ${\bf  -0.0010}$ & ${\bf  -0.7536}$ \\
$e$ & ${\it  0.0361}$ & ${\it  0.9998}$ & ${\bf  0.3778}$ & ${\bf  0.7540}$ & ${\bf  -0.7540}$ & ${\bf  -0.0010}$ & ${\bf  -0.3763}$ \\
$m_{\rm e}$ & ${\it  -0.0193}$ & ${\it  0.9993}$ & ${\it  0.9985}$ & ${\bf  1.5097}$ & ${\bf  -1.5097}$ & ${\bf  0.0011}$ & ${\bf  -0.7556}$ \\
$N_{\rm A}$ & ${\it  0.0193}$ & ${\it  -0.9993}$ & ${\it  -0.9985}$ & ${\it  -1.0000}$ & ${\bf  1.5097}$ & ${\bf  -0.0011}$ & ${\bf  0.7557}$ \\
$m_{\rm e}/m_{\rm \mu}$ & ${\it  -0.0202}$ & ${\it  -0.0004}$ & ${\it  -0.0007}$ & ${\it  0.0004}$ & ${\it  -0.0004}$ & ${\bf  4.9471}$ & ${\bf  -0.0021}$ \\
$F$ & ${\it  0.0745}$ & ${\it  -0.9957}$ & ${\it  -0.9939}$ & ${\it  -0.9985}$ & ${\it  0.9985}$ & ${\it  -0.0015}$ & ${\bf  0.3794}$ \\
\botrule
\end{tabular}
}
$^1$ The relative covariance is 
$u_{\rm r}(x_i,x_j) = u(x_i,x_j)/(x_ix_j)$, where $u(x_i,x_j)$ is the covariance of
$x_i$ and $x_j$; the relative variance is
$u_{\rm r}^2(x_i) = u_{\rm r}(x_i,x_i)$:
and the correlation coefficient is
$r(x_i,x_j) = u(x_i,x_j)/[u(x_i)u(x_j)]$.
\end{table*}

\begin{table*}
\caption{Internationally adopted values of various quantities.}
\tlabel{tab:adopted}
\def\hsp{\hbox to 22pt{}}
\begin{tabular}{l@{\hsp}l@{\hsp}l@{\hsp}l@{\hsp}l}
\toprule
& & & & Relative std. \\
~~~~~~Quantity & Symbol~~~~~ & Numerical value~~~~~ & Unit & uncert. $u_{\rm r}$ \\
\colrule
relative atomic mass$^1$ of $^{12}$C & $A_{\rm r}(^{12}$C) & $12$ &  & ~~~exact \vbox to 12 pt {} \\
molar mass constant  & $M_{\rm u}$ & $1 \times 10 ^{-3}$ & kg mol$^{-1}$ & ~~~exact \\
molar mass of $^{12}$C & $M(^{12}$C) & $12 \times 10 ^{-3}$ & kg mol$^{-1}$ & ~~~exact \\
conventional value of Josephson constant$^2$ & $K_{\rm J-90}$ & 483\,597.9 & GHz V$^{-1}$ & ~~~exact \\
conventional value of von Klitzing constant$^3$ ~~~~~ & $R_{\rm K-90}$ & 25\,812.807 & ${\rm \Omega}$ & ~~~exact \\
standard-state pressure & & $100$ & kPa & ~~~exact \\
standard atmosphere & & $101.325$ & kPa & ~~~exact \\
\botrule
\end{tabular}\\
$^1$ The relative atomic mass $A_{\rm r}(X)$ of particle $X$ with
 mass $m(X)$ is defined by $A_{\rm r}(X) = m(X) /m_{\rm u}$, where
$m_{\rm u} = m(^{12}{\rm C})/12 = M_{\rm u}/N_{\rm A} = 1~{\rm u}$ is the
atomic mass constant, $M_{\rm u}$ is the molar mass constant,
$N_{\rm A}$ is the Avogadro constant, and u is the unified
atomic mass unit.  Thus the mass of particle $X$ is $m(X) = A_{\rm r}(X)$~u
and the molar mass of $X$ is $M(X) = A_{\rm r}(X)M_{\rm u}$.
\\$^2$ This is the value adopted internationally
for realizing representations of the volt using the Josephson effect.
\\$^3$ This is the value adopted internationally
for realizing representations of the ohm using the quantum Hall effect.
\end{table*}

\begin{table*}[!]
\caption{Values of some x-ray-related quantities
based on the 2014 CODATA adjustment of the values of the constants.}
\tlabel{tab:xrayvalues}
\hbox to 7 in {
\begin{tabular}{llllll}
\toprule
& & & & Relative std. \\
~~~~~~Quantity & Symbol & ~~~~Numerical value & Unit & uncert. $u_{\rm r}$ \\
\colrule
Cu x unit: $\lambda({\rm CuK}{\rm \alpha}_{\rm 1}) / 1\,537.400 $ & ${\rm xu}({\rm CuK}{\rm \alpha}_{\rm 1})$ & $ 1.002\,076\,97(28)\times 10^{-13}$ & m & $ 2.8\times 10^{-7}$ 
\vbox to 12 pt {} \\
Mo x unit: $\lambda({\rm MoK}{\rm \alpha}_{\rm 1}) / 707.831 $ & ${\rm xu}({\rm MoK}{\rm \alpha}_{\rm 1})$ & $ 1.002\,099\,52(53)\times 10^{-13}$ & m & $ 5.3\times 10^{-7}$ \\
{\aa}ngstrom star$: \lambda({\rm WK}{\rm \alpha}_{\rm 1}) / 0.209\,010\,0 $ & \AA$^{\ast}$ & $ 1.000\,014\,95(90)\times 10^{-10}$ & m & $ 9.0\times 10^{-7}$ \\
lattice parameter$^1$ of Si \ (in vacuum, 22.5 $^\circ$C)~~ & $a$ & $ 543.102\,0504(89)\times 10^{-12}$ & m & $ 1.6\times 10^{-8}$ \vbox to 9 pt {}\\
\{220\} lattice spacing of Si $a/\sqrt{8}$ & $d_{\rm 220}$ & $ 192.015\,5714(32)\times 10^{-12}$ & m & $ 1.6\times 10^{-8}$ \\
\,\, (in vacuum, 22.5 $^\circ$C)&&&&\\
molar volume of Si \ $M({\rm Si})/\rho({\rm Si})=N_{\rm A}a^{3}\!/8$ & $V_{\rm m}$(Si) & $ 12.058\,832\,14(61)\times 10^{-6}$ & m$^{3}$ mol$^{-1}$ & $ 5.1\times 10^{-8}$ \\
\, \, (in vacuum, 22.5 $^\circ$C)&&&&\\
\botrule
\end{tabular}
}
$^1$ This is the lattice parameter (unit cell edge length) 
of an ideal single crystal of naturally occurring
Si free of impurities and imperfections, and is deduced from
measurements on extremely pure and nearly perfect single crystals of Si
by correcting for the effects of impurities.
\end{table*}

\thispagestyle{empty}
\def\hsp{\hbox to 15 pt {}}
\begin{table*}[!]
\caption{The values in SI units of some non-SI units based on the 2014
CODATA adjustment of the values of the constants.}
\tlabel{tab:units}
\begin{tabular}{l@{\hsp}l@{\hsp}l@{\hsp}l@{\hsp}l@{\hsp}l}
\toprule
& & & & Relative std. \\
\s{35}Quantity & \s{-3}Symbol & \s{17}Numerical value & \s{2}Unit & uncert. $u_{\rm r}$ \\
\colrule
\multicolumn {5} {c} { \vbox to 12 pt {} Non-SI units accepted for use with the SI} \\
electron volt: ($e/{\rm C}$) {\rm J} & eV & $ 1.602\,176\,6208(98)\times 10^{-19}$ & J & $ 6.1\times 10^{-9}$ \\
(unified) atomic mass unit: ${1\over12}m(^{12}$C)~~ & u & $ 1.660\,539\,040(20)\times 10^{-27}$ & kg & $ 1.2\times 10^{-8}$ \\
\multicolumn {5} {c} {} \\
\multicolumn {5} {c} {Natural units (n.u.)} \\
n.u. of velocity & $c,c_0$ & 299\,792\,458 & m s$^{-1}$ & exact \\
n.u. of action: $h/2\rmpi$~~ & $\hbar$ & $ 1.054\,571\,800(13)\times 10^{-34}$ & J s & $ 1.2\times 10^{-8}$ \\
& & $ 6.582\,119\,514(40)\times 10^{-16}$ & eV s & $ 6.1\times 10^{-9}$ \\
& $\hbar c$ & $ 197.326\,9788(12)$ & MeV fm & $ 6.1\times 10^{-9}$ \\
n.u. of mass & $m_{\rm e}$ & $ 9.109\,383\,56(11)\times 10^{-31}$ & kg & $ 1.2\times 10^{-8}$ \\
n.u. of energy & $m_{\rm e}c^2$ & $ 8.187\,105\,65(10)\times 10^{-14}$ & J & $ 1.2\times 10^{-8}$ \\
& & $ 0.510\,998\,9461(31)$ & MeV & $ 6.2\times 10^{-9}$ \\
n.u. of momentum & $m_{\rm e}c$ & $ 2.730\,924\,488(34)\times 10^{-22}$ & kg m s$^{-1}$ & $ 1.2\times 10^{-8}$ \\
& & $ 0.510\,998\,9461(31)$ & MeV/$c$ & $ 6.2\times 10^{-9}$ \\
n.u. of length: $\hbar/m_{\rm e}c$& $\lbar_{\rm C}$ & $ 386.159\,267\,64(18)\times 10^{-15}$ & m & $ 4.5\times 10^{-10}$ \\
n.u. of time & $\hbar/m_{\rm e}c^2$ & $ 1.288\,088\,667\,12(58)\times 10^{-21}$ & s & $ 4.5\times 10^{-10}$ \\
\multicolumn {5} {c} {} \\
\multicolumn {5} {c} {Atomic units (a.u.)} \\
a.u. of charge & $e$ & $ 1.602\,176\,6208(98)\times 10^{-19}$ & C & $ 6.1\times 10^{-9}$ \\
a.u. of mass & $m_{\rm e}$ & $ 9.109\,383\,56(11)\times 10^{-31}$ & kg & $ 1.2\times 10^{-8}$ \\
a.u. of action: $h/2\rmpi$& $\hbar$ & $ 1.054\,571\,800(13)\times 10^{-34}$ & J s & $ 1.2\times 10^{-8}$ \\
a.u. of length: Bohr radius (bohr)~~& & & & \\
\, $\alpha/4\rmpi R_\infty$& $a_0$ & $ 0.529\,177\,210\,67(12)\times 10^{-10}$ & m & $ 2.3\times 10^{-10}$ \\
a.u. of energy: Hartree energy (hartree)~~ & & & & \\
\, $e^2\!/4\rmpi\epsilon_0a_0=2R_\infty hc = \alpha^2m_{\rm e}c^2$ & $E_{\rm h}$ & $ 4.359\,744\,650(54)\times 10^{-18}$ & J & $ 1.2\times 10^{-8}$ \\
a.u. of time & $\hbar/E_{\rm h}$ & $ 2.418\,884\,326\,509(14)\times 10^{-17}$ & s & $ 5.9\times 10^{-12}$ \\
a.u. of force & $E_{\rm h}/a_0$ & $ 8.238\,723\,36(10)\times 10^{-8}$ & N & $ 1.2\times 10^{-8}$ \\
a.u. of velocity: $\alpha c$ & $a_0E_{\rm h}/\hbar$ & $ 2.187\,691\,262\,77(50)\times 10^{6}$ & m s$^{-1}$ & $ 2.3\times 10^{-10}$ \\
a.u. of momentum & $\hbar/a_0$ & $ 1.992\,851\,882(24)\times 10^{-24}$ & kg m s$^{-1}$ & $ 1.2\times 10^{-8}$ \\
a.u. of current & $eE_{\rm h}/\hbar$ & $ 6.623\,618\,183(41)\times 10^{-3}$ & A & $ 6.1\times 10^{-9}$ \\
a.u. of charge density & $e/a_0^3$ & $ 1.081\,202\,3770(67)\times 10^{12}$ & C m$^{-3}$ & $ 6.2\times 10^{-9}$ \\
a.u. of electric potential & $E_{\rm h}/e$ & $ 27.211\,386\,02(17)$ & V & $ 6.1\times 10^{-9}$ \\
a.u. of electric field & $E_{\rm h}/ea_0$ & $ 5.142\,206\,707(32)\times 10^{11}$ & V m$^{-1}$ & $ 6.1\times 10^{-9}$ \\
a.u. of electric field gradient & $E_{\rm h}/ea_0^2$ & $ 9.717\,362\,356(60)\times 10^{21}$ & V m$^{-2}$ & $ 6.2\times 10^{-9}$ \\
a.u. of electric dipole moment & $ea_0$ & $ 8.478\,353\,552(52)\times 10^{-30}$ & C m & $ 6.2\times 10^{-9}$ \\
a.u. of electric quadrupole moment & $ea_0^2$ & $ 4.486\,551\,484(28)\times 10^{-40}$ & C m$^2$ & $ 6.2\times 10^{-9}$ \\
a.u. of electric polarizability & $e^2a_0^2/E_{\rm h}$ & $ 1.648\,777\,2731(11)\times 10^{-41}$ & C$^2$ m$^2$ J$^{-1}$ & $ 6.8\times 10^{-10}$ \\
a.u. of 1$^{\rm st}$ hyperpolarizability & $e^3a_0^3/E_{\rm h}^2$ & $ 3.206\,361\,329(20)\times 10^{-53}$ & C$^3$ m$^3$ J$^{-2}$ & $ 6.2\times 10^{-9}$ \\
a.u. of 2$^{\rm nd}$ hyperpolarizability & $e^4a_0^4/E_{\rm h}^3$ & $ 6.235\,380\,085(77)\times 10^{-65}$ & C$^4$ m$^4$ J$^{-3}$ & $ 1.2\times 10^{-8}$ \\
a.u. of magnetic flux density & $\hbar/ea_0^2$ & $ 2.350\,517\,550(14)\times 10^{5}$ & T & $ 6.2\times 10^{-9}$ \\
a.u. of magnetic dipole moment: $2\mu_{\rm B}$ & $\hbar e/m_{\rm e}$ & $ 1.854\,801\,999(11)\times 10^{-23}$ & J T$^{-1}$ & $ 6.2\times 10^{-9}$ \\
a.u. of magnetizability & $e^2a_0^2/m_{\rm e}$ & $ 7.891\,036\,5886(90)\times 10^{-29}$ & J T$^{-2}$ & $ 1.1\times 10^{-9}$ \\
a.u. of permittivity: $10^7/c^2$ & $e^2/a_0E_{\rm h}$ & $ 1.112\,650\,056\ldots\times 10^{-10}$ & F m$^{-1}$ & exact \\
\botrule
\end{tabular}
\end{table*}

\begingroup
\squeezetable
\begin{table*}[!]
\def\hsp{\hbox to 16 pt {}}
\caption{The values of some energy equivalents derived from the
relations $E=mc^2 = hc/\lambda = h\nu = kT$, and based on the 2010
CODATA adjustment of the values of the constants; 1~eV~$=(e/{\rm C})$~J,
1~u $= m_{\rm u} = \textstyle{1\over12}m(^{12}{\rm C}) =
10^{-3}$~kg~mol$^{-1}\!/N_{\rm A}$, and $E_{\rm h} = 2R_{\rm \infty}hc =
\alpha^2m_{\rm e}c^2$ is the Hartree energy (hartree).}
\tlabel{tab:enconv1}
\begin{tabular} {l@{\hsp}l@{\hsp}l@{\hsp}l@{\hsp}l}
\toprule
\multicolumn{5} {c} {Relevant unit} \\
\colrule
    &   \s{30}J &  \s{30}kg &  \s{30}m$^{-1}$  &   \s{30}Hz  \vbox to 12 pt {}   \\
\colrule
        &                    &                    &               &            \\
1~J    & $(1\ {\rm J})=$    &  (1 J)/$c^2=$     &       (1 J)/$hc=$   &        (1 J)/$h=$              \\
 & 1 J   & $ 1.112\,650\,056\ldots\times 10^{-17}$ kg  & $ 5.034\,116\,651(62)\times 10^{24}$ m$^{-1}$ & $ 1.509\,190\,205(19)\times 10^{33}$ Hz \\
        &        &             &           &  \\
1~kg    &        (1 kg)$c^2=$   &   $(1 \ {\rm kg})=$       &        (1 kg)$c/h=$ &        (1 kg)$c^2/h=$          \\
 &  $ 8.987\,551\,787\ldots\times 10^{16}$ J & 1 kg   & $ 4.524\,438\,411(56)\times 10^{41}$ m$^{-1}$ & $ 1.356\,392\,512(17)\times 10^{50}$ Hz \\
 &           &            &         &   \\
1~m$^{-1}$   &  (1 m$^{-1})hc=$   &       (1 m$^{-1})h/c=$    &  $(1$ m$^{-1})=$    &   (1 m$^{-1})c=$  \\
 &  $ 1.986\,445\,824(24)\times 10^{-25}$ J  & $ 2.210\,219\,057(27)\times 10^{-42}$ kg  & 1 m$^{-1}$  & $ 299\,792\,458$ Hz \\
 &           &            &         &   \\
1~Hz   &  (1 Hz)$h=$  &  (1 Hz)$h/c^{2}=$   &        (1 Hz)/$c=$  &  $(1$ Hz$)=$  \\
 &  $ 6.626\,070\,040(81)\times 10^{-34}$ J  & $ 7.372\,497\,201(91)\times 10^{-51}$ kg & $ 3.335\,640\,951\ldots\times 10^{-9}$ m$^{-1}$ & 1 Hz \\
 &           &            &         &   \\
1~K  &  (1 K)$k=$  &   (1 K)$k/c^{2}=$  &    (1 K)$k/hc=$ &  (1 K)$k/h=$   \\
 &  $ 1.380\,648\,52(79)\times 10^{-23}$ J  & $ 1.536\,178\,65(88)\times 10^{-40}$ kg & $ 69.503\,457(40)$ m$^{-1}$ & $ 2.083\,6612(12)\times 10^{10}$ Hz \\
         &            &            &           &        \\
1~eV    &  (1 eV) =  &  $(1~{\rm eV})/c^{2}=$  &   $(1~{\rm eV})/hc=$   &     $(1~{\rm eV})/h=$   \\
 &  $ 1.602\,176\,6208(98)\times 10^{-19}$ J  & $ 1.782\,661\,907(11)\times 10^{-36}$ kg & $ 8.065\,544\,005(50)\times 10^{5}$ m$^{-1}$ & $ 2.417\,989\,262(15)\times 10^{14}$ Hz \\
         &          &             &         &     \\
1~u   &   $(1~{\rm u})c^{2}=$   &  (1 u) =   &  $(1~{\rm u})c/h=$  &  $(1~{\rm u})c^{2}/h=$    \\
 &  $ 1.492\,418\,062(18)\times 10^{-10}$ J  & $ 1.660\,539\,040(20)\times 10^{-27}$ kg & $ 7.513\,006\,6166(34)\times 10^{14}$ m$^{-1}$ & $ 2.252\,342\,7206(10)\times 10^{23}$ Hz \\
         &          &             &        &       \\
1~$E_{\rm h}$   &  $(1~E_{\rm h})=$  & $(1~E_{\rm h})/c^2=$ &  $(1~E_{\rm h})/hc=$  &  $(1~E_{\rm h})/h=$   \\
 &  $ 4.359\,744\,650(54)\times 10^{-18}$ J  & $ 4.850\,870\,129(60)\times 10^{-35}$ kg & $ 2.194\,746\,313\,702(13)\times 10^{7}$ m$^{-1}$ & $ 6.579\,683\,920\,711(39)\times 10^{15}$ Hz \\
\botrule
\end{tabular}
\end{table*}
\endgroup

\begingroup
\squeezetable
\begin{table*}[!]
\def\hsp{\hbox to 16 pt {}}
\caption{The values of some energy equivalents derived from the
relations $E=mc^2 = hc/\lambda = h\nu = kT$, and based on the 2010
CODATA adjustment of the values of the constants; 1~eV~$=(e/{\rm C})$~J,
1~u $= m_{\rm u} = \textstyle{1\over12}m(^{12}{\rm C}) =
10^{-3}$~kg~mol$^{-1}\!/N_{\rm A}$, and $E_{\rm h} = 2R_{\rm \infty}hc =
\alpha^2m_{\rm e}c^2$ is the Hartree energy (hartree).}
\tlabel{tab:enconv2}
\begin{tabular} {l@{\hsp}l@{\hsp}l@{\hsp}l@{\hsp}l}
\toprule
\multicolumn{5} {c} {Relevant unit} \\
\colrule
    &   \s{30}K &  \s{30}eV &  \s{30}u  &   \s{30}$E_{\rm h}$  \vbox to 12 pt {}   \\
\colrule
        &                    &                    &               &            \\
1~J     & (1 J)/$k=$   &  (1 J) =   &       (1 J)/$c^2$ = &   (1 J) =              \\
 &  $ 7.242\,9731(42)\times 10^{22}$ K  & $ 6.241\,509\,126(38)\times 10^{18}$ eV  & $ 6.700\,535\,363(82)\times 10^{9}$ u & $ 2.293\,712\,317(28)\times 10^{17}$ $E_{\rm h}$ \\
        &        &             &           &  \\
1~kg    &        (1 kg)$c^2/k=$    &   (1 kg)$c^2$ =    &    (1 kg) =  &  (1 kg)$c^2=$   \\
 &  $ 6.509\,6595(37)\times 10^{39}$ K  & $ 5.609\,588\,650(34)\times 10^{35}$ eV  & $ 6.022\,140\,857(74)\times 10^{26}$ u  & $ 2.061\,485\,823(25)\times 10^{34}$ $E_{\rm h}$ \\
 &           &            &         &   \\
1~m$^{-1}$   &  (1 m$^{-1})hc/k=$  &       (1 m$^{-1})hc=$  &  (1 m$^{-1})h/c$ =   &   (1 m$^{-1})hc=$ \\
 &  $ 1.438\,777\,36(83)\times 10^{-2}$ K  & $ 1.239\,841\,9739(76)\times 10^{-6}$ eV  & $ 1.331\,025\,049\,00(61)\times 10^{-15}$ u & $ 4.556\,335\,252\,767(27)\times 10^{-8}$ $E_{\rm h}$ \\
         &           &            &            &      \\
1~Hz   &  (1 Hz)$h/k=$  &  (1 Hz)$h=$  &   (1 Hz)$h/c^2$ = & (1 Hz)$h=$  \\
 &  $ 4.799\,2447(28)\times 10^{-11}$ K  & $ 4.135\,667\,662(25)\times 10^{-15}$ eV  & $ 4.439\,821\,6616(20)\times 10^{-24}$ u & $ 1.519\,829\,846\,0088(90)\times 10^{-16}$ $E_{\rm h}$ \\
           &             &          &           &        \\ 
1~K  & $(1$ K$)=$  &   (1 K)$k=$ &    (1 K)$k/c^2=$ &  (1 K)$k=$   \\
 & 1 K   &   $ 8.617\,3303(50)\times 10^{-5}$ eV & $ 9.251\,0842(53)\times 10^{-14}$ u & $ 3.166\,8105(18)\times 10^{-6}$ $E_{\rm h}$ \\
         &            &            &           &        \\
1~eV    &  (1 eV)/$k=$  &  $(1$ eV$)=$  &   $(1~{\rm eV})/c^2=$  &     $(1~{\rm eV})=$   \\
 &  $ 1.160\,452\,21(67)\times 10^{4}$ K  & 1 eV  & $ 1.073\,544\,1105(66)\times 10^{-9}$ u & $ 3.674\,932\,248(23)\times 10^{-2}$ $E_{\rm h}$ \\
         &          &             &         &     \\
1~u   &   $(1~{\rm u})c^{2}/k=$   &  $(1~{\rm u})c^2=$ &  $(1$ u$)=$  &  $(1~{\rm u})c^2=$   \\
 &  $ 1.080\,954\,38(62)\times 10^{13}$ K  & $ 931.494\,0954(57)\times 10^{6}$ eV  & 1 u & $ 3.423\,177\,6902(16)\times 10^{7}$ $E_{\rm h}$ \\
         &          &             &        &       \\
1~$E_{\rm h}$   &  $(1~E_{\rm h})/k=$  & $(1~E_{\rm h})=$ &  $(1~E_{\rm h})/c^2=$ & $(1~E_{\rm h})=$    \\
 &  $ 3.157\,7513(18)\times 10^{5}$ K  & $ 27.211\,386\,02(17)$ eV  & $ 2.921\,262\,3197(13)\times 10^{-8}$ u  & $1~E_{\rm h}$  \\
\botrule
\end{tabular}
\end{table*}
\endgroup

\end{document}